# Aligning geographic entities from historical maps for building knowledge graphs

Kai Sun[a,b,c], Yingjie Hu[c], Jia Song[a,d], Yunqiang Zhu[a,d]

a. State Key Laboratory of Resources and Environmental Information System, Institute of Geographic Sciences and Natural Resources Research, Chinese Academy of Sciences, Beijing, China;
b. University of Chinese Academy of Sciences, Beijing, China;
c. GeoAI Lab, Department of Geography, University at Buffalo, Buffalo, NY, USA;
d. Jiangsu Center for Collaborative Innovation in Geographical Information Resource Development and Application, Nanjing, China

**Abstract:** Historical maps contain rich geographic information about the past of a region. They are sometimes the only source of information before the availability of digital maps. Despite their valuable content, it is often challenging to access and use the information in historical maps, due to their forms of paper-based maps or scanned images. It is even more time-consuming and labor-intensive to conduct an analysis that requires a synthesis of the information from multiple historical maps. To facilitate the use of the geographic information contained in historical maps, one way is to build a geographic knowledge graph (GKG) from them. This paper proposes a general workflow for completing one important step of building such a GKG, namely aligning the same geographic entities from different maps. We present this workflow and the related methods for implementation, and systematically evaluate their performances using two different datasets of historical maps. The evaluation results show that machine learning and deep learning models for matching place names are sensitive to the thresholds learned from the training data, and a combination of measures based on string similarity, spatial distance, and approximate topological relation achieves the best performance with an average F-score of 0.89.

## 1 Introduction

For centuries, maps have been the major medium for depicting and communicating geographic features and their locations. Historical maps are unique and valuable archives that document the geographic features in different time periods, and are sometimes the only source of information for analyzing the change of a geographic area over time (Chiang et al. 2014, Uhl et al. 2018). Fortunately, many historical maps are scanned and preserved as digital images, such as the historical topographic maps provided by the USGS[1], the Sanborn Fire Insurance maps[2], and the many map collections preserved in local libraries. These historical maps carry important geographic information, such as the names and locations of natural and man-made features (Kaim et al. 2016, Liu et al. 2018), and are critical for understanding the past of a geographic region and for answering various location-related questions.

---
[1] http://historicalmaps.arcgis.com/usgs/

While many historical maps have been scanned into images, it is often difficult to find and use the geographic information contained in these map images. For example, to see the historical buildings and streets of a neighborhood in a particular year, one needs to first find the relevant historical maps through e.g., a library search system. Since a large map is often scanned into multiple images each of which covers a small fragment of the map, one also has to understand the image indexing method or has to manually browse many map images to find the relevant ones. When it comes to analyzing the changes during multiple time periods, one has to repeat the above process multiple times to identify the corresponding maps, and needs to manually synthesize the information from these maps to see the changes. Due to these difficulties, it is often time-consuming and labor-intensive to use the information from only a few historical maps, let alone conducting a large-scale analysis over many maps.

Geographic knowledge graph (GKG) provides one promising approach for facilitating the access and use of the geographic information contained in historical maps (Chiang 2015, Lin et al. 2018, Varanka and Usery 2018). A knowledge graph is an organized collection of entities and concepts linked via their possible relations (Singhal 2012). A GKG can be considered as a special type of knowledge graph that focuses on entities and concepts in the domain of geography (Ballatore et al. 2015, Yan et al. 2019, Wang et al. 2019). Similar to a knowledge graph, a GKG can support various queries especially those focusing on geographic entities. The Linked Open Data (LOD) cloud already provides some GKGs, such as GeoNames and LinkedGeoData (Auer et al. 2009). Some studies also developed their own GKGs, such as a GKG for geoscience literature (Wang et al. 2018), a GKG for urban computing (Zhang et al. 2018), and a GKG for crowdsourced geographic information from OpenStreetMap and Wikidata (Chen et al. 2017).

Given a collection of historical maps, a GKG can be constructed by first extracting geographic entities from the scanned map images and then aligning the corresponding entities from different maps (e.g., via *same-as* relations). With such a GKG, we can quickly answer queries such as "show me the buildings along Washington Street that existed in 1925 but were not there in 1889". By contrast, one traditionally may have to manually identify these buildings from the two historical maps in 1889 and 1925 respectively, and then make a comparison among the identified buildings.

The topic of extracting vector entities from scanned raster map images has attracted the attention of researchers for decades (Freeman and Pieroni 1980). Many methods have been developed for extracting different types of entities and information from map images. Examples include the methods for extracting textual labels (Leyk et al. 2006, Pezeshk and Tutwiler 2011, Xu et al. 2016), road segments (Callier and Saito 2012, Chiang and Knoblock 2013), road intersections (Chiang et al. 2009, Saeedimoghaddam and Stepinski 2019), contour lines (Liu et al. 2016, Song et al. 2016), built-up areas (Muhs et al. 2016), and map symbols (Miao et al. 2016). In addition, there also exist software tools, such as ArcScan from ArcGIS and Polygonize from QGIS, that can be used for extracting vector entities from scanned map images. On the other hand, it has been shown that different map collections often have their own colors, hues, contrasts, and cartographic styles (Uhl et al. 2018). As a result, there is no one single extraction method that can be effectively applied to all different historical maps. Given a particular map collection, one may have to try multiple methods in order to find the most effective ones for extracting entities from map images.

---
[2] https://www.loc.gov/collections/sanborn-maps

In this work, we assume that geographic entities have already been extracted using selected methods, and we focus on the task of identifying and linking the same entities from different maps. Our goal is to develop an automatic entity alignment workflow, which can be connected to the image-based entity extraction methods selected by a user to convert a collection of historical maps into a GKG. With such a goal, aligning the geographic entities extracted from different historical maps faces two major challenges. First, many historical maps, especially those from local libraries, are simply scanned images that are not georeferenced. Consequently, the entities extracted from these map images have no georeference either. Since the maps can be in different orientations, covering different spatial extents, and in different geographic scales, the extracted entities cannot be directly overlaid to identify alignments. Second, historical maps may provide only partial information for some entities. For example, Figure 1 shows a fragment of a digital Sanborn Fire Insurance map for the city of Albany, NY, USA in the year of 1892. As can be seen, many city blocks are not associated with place names or any textual labels, and the extracted city block entities will not have any textual labels either. Due to these two challenges, aligning the geographic entities from different historical maps is prone to both false positives (i.e., when different entities are incorrectly aligned) and false negatives (i.e., when corresponding entities are not aligned). This paper aims to address these challenges and our contributions are as follows:
- We propose a general workflow for automatically aligning geographic entities extracted from different historical maps.
- We discuss the possible methods for the individual steps of the proposed workflow, and evaluate their performances on two datasets of historical maps.
- We share the source code of our entity alignment workflow and the two test datasets of historical maps to support future research.

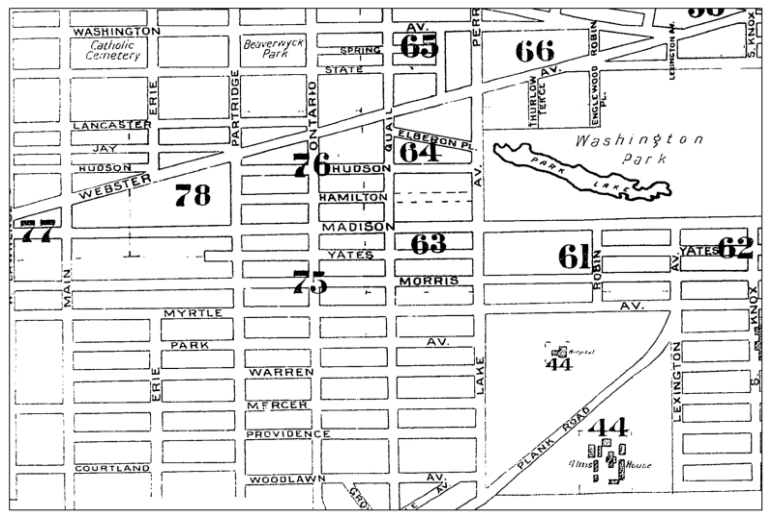

Figure 1. A fragment of a historical map for the city of Albany, NY, USA in 1892.

The remainder of this paper is organized as follows. Section 2 provides a review of related work on geospatial data matching, map conflation, and geographic entity alignment. Section 3 presents our general workflow for aligning entities from different historical maps and discusses the possible methods for implementation. In Section 4, we evaluate the proposed workflow and the related methods by applying them to two different datasets of historical maps. We present the evaluation results and discuss the advantages and limitations of the proposed workflow and methods. Finally, Section 5 summarizes this work and discusses future directions.

## 2 Related work
### 2.1 Geospatial data matching and conflation

Two topics closely related to this work are geospatial data matching and map conflation. Geospatial data matching refers to the process of identifying corresponding data records that represent the same real-world entities from two geospatial datasets (Walter and Fritsch 1999, Xavier et al. 2016). Map conflation, or more generally geospatial data conflation, makes one step further by conflating the matching data records from different geospatial datasets in order to generate a better and richer dataset (Lynch and Saalfeld 1985, Ruiz et al. 2011).

Existing research on geospatial data matching and conflation often involves two aspects: similarity measures and matching methods. Similarity measures quantify the similarity between two data records (e.g., two geographic features), and matching methods utilize one or multiple similarity measures to determine the way in which two data records should be matched. Similarity measures frequently used in the literature include geometric measures, attribute measures, and topological measures (Tong et al. 2009, Zhang et al. 2012, Xavier et al. 2016). Geometric measures focus on the geometric properties of geospatial features, such as position, size, and shape, and often leverage spatial distance (Yuan and Tao 1999, Devogele 2002, Deng et al. 2007), overlapping area (van Wijngaarden et al. 1997, Hastings 2008, Fan et al. 2016), and others to quantify the similarity between two features. Attribute measures focus on the non-geometric properties of geospatial features. Depending on the data type of the attribute, different measures may be adopted, such as scalar similarity for numeric attributes (Samal et al. 2004) and string similarity or phonetic similarity for textual attributes (Li and Goodchild 2011, McKenzie et al. 2014). Attribute measures are also considered as semantic measures (Devogele et al. 1996, Balley et al. 2004, Zhang 2009), since they often need to quantify the similarity of different concepts or terms based on their meanings. Knowledge networks and ontologies are often used for quantifying semantic similarity (Rada et al. 1989, Fonseca et al. 2002, Schwering 2008). Topological measures focus on the topological and spatial relations among geospatial features (Egenhofer et al. 1993, Filin and Doytsher 2000, Hope and Kealy 2008). Existing research has proposed and used various measures based on topological relations, such as *node degree* (Rosen and Saalfeld 1985), *connected* (Walter and Fritsch 1999), and *neighborhood* (Olteanu-Raimond and Mustière 2008), to quantify the similarity between geospatial features. Topological relations have also been used to define geographic contexts and measure their similarity (Samal et al. 2004).

There exist many methods for geospatial data matching and conflation, which can be categorized from different views. Depending on the number of similarity measures used for matching, there are single-measure-based methods and multiple-measure-based ones. When multiple similarity measures are used, they can be combined via approaches, such as a weighted combination (Zhang and Meng 2007), an optimization model (Li and Goodchild 2011), and the belief theory (Olteanu-Raimond et al. 2015). Machine learning models, such as support vector machines (Martins 2011), artificial neural networks (Li et al. 2012), and random forests (Acheson et al. 2019), are also used for combining multiple measures. Depending on the main type of similarity measures used for matching, there are geometric, attribute, and topological matching methods (Yuan and Tao 1999). Depending on the target to be matched, there are schema, feature, and internal matching methods (Xavier et al. 2016). While schema and feature level methods focus on matching the concepts and instances from two datasets respectively, internal methods are applied to the interior components of features (e.g., vertices). Depending on the geometry type of the geographic features to be matched, there are point-based, line-based, area-based, and mixed methods (Volz 2006). Depending on the data model of the geographic features to be matched, there are vector-vector, raster-raster, and vector-raster methods (Chen et al. 2008, Seo and O'Hara 2009). Along this direction, researchers also developed spatial data models that link entities across vector and raster layers, such as the hierarchical spatial model (Yuan 2001), object field (Cova and Goodchild 2002), and geo-atom (Goodchild et al. 2007). Depending on the number of cases to be matched between geographic features, there are one-to-one, one-to-many, and many-to-many matching methods (Beeri et al. 2004, Mustière and Devogele 2008). Many other studies exist in the area of geospatial data matching, and interested readers can refer to the review by Xavier et al. (2016).

## 2.2 Geospatial entity alignment

The task of *entity alignment* is very similar to data matching but is typically discussed under the context of knowledge graphs (Sun et al. 2018, Trisedya et al. 2019, Pei et al. 2019). Geographic entity alignment focuses on matching the entities that have a geographic meaning, such as the geographic features on maps. It is worth noting that the term *alignment* is often used in GIScience literature to refer to the process of re-positioning the geometry of a geographic feature by adjusting the coordinates of its vertices. Geographic entity alignment, however, does not adjust the geometry of an entity but matches or links the entities that represent the same real-world geographic objects. Like geospatial data matching, geographic entity alignment also involves the two aspects of similarity measures and matching methods. Sun et al. (2019) considered three steps for geographic entity alignment, which are property selection, similarity calculation, and alignment classification. The first step selects the properties of the entities to be compared; the second step computes similarity scores based on the selected properties and similarity measures; and the third step uses a method to combine the similarity scores to determine the alignment result. Many of the similarity measures and matching methods used for geospatial data matching, such as those based on geometries, attributes, and topological relations, are also used for geographic entity alignment (Sehgal et al. 2006, Yu et al. 2018).

Compared with research in geospatial data matching and entity alignment, aligning entities from historical maps often needs to handle the challenges of partially available information and the lack of georeferences. Existing studies were often done on datasets with comprehensive information about the entities to be matched or aligned. For example, Santos et al. (2018a) and Santos et al. (2018b) matched the entities from the GeoNames gazetteer in which every entity has information about their official and alternative place names. Yu et al. (2018) aligned the entities from GeoNames and OpenStreetMap, in which the entities have comprehensive information about their place names, place types, spatial footprints, and hierarchical relations of these entities. There also exist many other studies that aligned entities from the data sources in which comprehensive information about the entities are available (Auer et al. 2009, Stadler et al. 2012, McKenzie et al. 2014). By contrast, the geographic entities from historical maps often have only partial information. Many entities may not have place names or place types, and this can directly limit the applicability of some matching methods that rely on the string similarity of place names or place type similarity.

In addition, many historical map images do not have georeferences, and consequently, the similarity measures and matching methods based on geometries and spatial coordinates cannot be directly used. In the map conflation literature, researchers used manual or semi-automatic rubber sheeting to match and conflate non-georeferenced map data (White Jr and Griffin 1985, Lupien and Moreland 1987, Shimizu and Fuse 2003). Rubber sheeting is a method for spatially adjusting one map so that it can be overlaid on another (Saalfeld 1988). For historical maps without georeferencing information, rubber sheeting can be applied to adjusting the positions of the extracted entities. Besides, topological relations, which are usually invariant to map orientations and scales, can be used for aligning entities from historical maps as well.

Similar to geospatial data matching, the result of geographic entity alignment can be in the forms of one-to-one, one-to-many, and many-to-many. In this work, we focus on one-to-one alignment only. This is different from some studies that include the other types of alignments. For example, in aligning road networks with different levels of details, Mustière and Devogele (2008) allowed one road from the coarse dataset to be aligned to multiple roads in the more-detailed dataset (i.e., one-to-many alignment). In this work, while one geographic feature may be represented as one or multiple entities on different maps, we do not attempt to align them. This is because it is very difficult to differentiate these representation differences from the actual changes in the real world when historical maps are examined. For example, one city block from an earlier map may indeed be separated into two blocks by a newly developed street in a latter map. Thus, directly aligning those entities could lead to many false positives, and manual examination on these not-aligned entities may be a more suitable way for determining their relations. From this perspective, this work is similar to the studies, such as the one from Beeri et al. (2004), that focus on one-to-one matching.

## 3 Methods
This section presents the methodological details of our proposed workflow for aligning geographic entities from historical maps. We start by formalizing the problem, then provide an overview of the workflow, and finally discuss the possible methods for implementing the individual steps of the workflow.

### 3.1 Problem formalization
In this alignment problem, geographic entities have been extracted from scanned historical map images as vector data (i.e., points, lines, and polygons), and the objective is to align the entities from different maps in order to construct a GKG. The entities are accompanied with only the information available on the maps. For example, a street will not have a place name if its name is not provided on the map. Similarly, the extracted entities will not have correct geographic coordinates if the georeferencing information is not available (note that the entities will still have relative coordinates based on their relative positions on the map image). We formalize the problem addressed in this work as below:

*Given the geographic entities from historical map A, $\{e_{A1}, e_{A2}, e_{A3}, …, e_{An}\}$, and the geographic entities from historical map B, $\{e_{B1}, e_{B2}, e_{B3}, …, e_{Bm}\}$, identify matching entity pairs, $\{[e_{Ax1}, e_{By1}], [e_{Ax2}, e_{By2}], … \}$, that represent the same real-world features.*

With this problem formalization, the alignment is performed between entities from two maps. For a collection of three or more maps, we can perform alignment between the maps in any two consecutive timestamps and combine the result. For example, for a set of three maps $\{map_1, map_2, map_3\}$, we will perform alignment between $map_1$ and $map_2$ and between $map_2$ and $map_3$ respectively, assuming $time_{map1} < time_{map2} < time_{map3}$. In addition, we focus on one-to-one alignment only in this work, and one entity from the first map can be aligned to at most one entity in the second map. It is possible that an entity may not have a corresponding entity in the other map, and thus will not be aligned to any entity. In the following, we present the workflow and its individual steps.

## 3.2 Overall workflow

We develop a general workflow for aligning geographic entities from historical maps. Figure 2 provides an overview of this workflow.

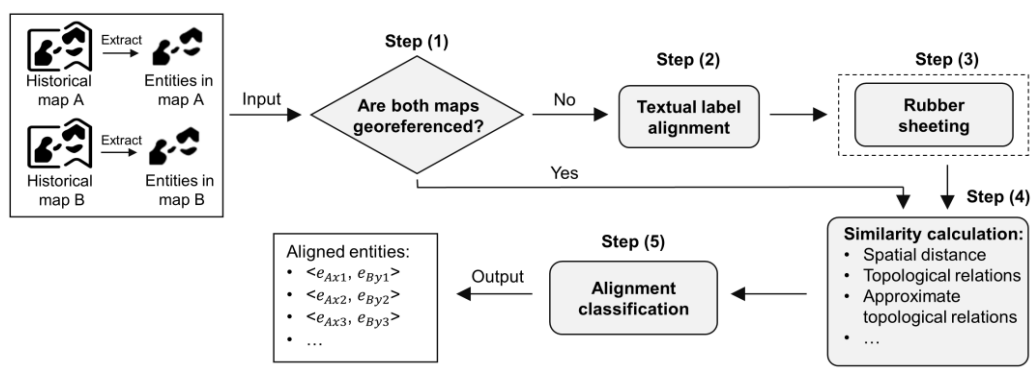

Figure 2. A general workflow for aligning geographic entities from historical maps.

The input of the workflow contains the entities extracted from historical maps. Extraction methods from existing research, such as those from Chiang and Knoblock (2013), Pezeshk and Tutwiler (2011), and Muhs et al. (2016), and software tools, such as ArcScan, can be employed to extract vector entities and textual labels from raster maps. The entities will need to have three basic properties before they are passed to the workflow as the input, which are place name, geometric footprint, and unique ID. The first two are extracted from the historical maps, and some entities may not have place names if they are not available on the maps. The unique ID is simply a string for identifying each entity, and can be automatically generated during the entity extraction process.

With the input entities prepared, step (1) of the workflow examines whether the two historical maps have proper georeferences. If luckily "Yes", it overlays the entities from these two maps (after performing necessary map projection transformations), and proceeds to step (4) which computes the spatial distances, topological relations, and other measurements among the entities and calculates their corresponding similarity scores. A more challenging situation is when either (or both) of the maps does not have georeference. Since the maps can be in different orientations and covering different geographic extents, the entities cannot be simply overlaid. In such a situation, the workflow moves to step (2) and performs *textual label* *alignment* to identify the entities that can be aligned based on their names, e.g., "Delaware Street" on map A may be aligned to "Delaware St." on map B. With the initial set of entities aligned based on their textual labels, the workflow automatically evaluates whether *rubber sheeting* can be performed to further adjust the spatial positions of the entities so that they can be put under the same spatial reference. If so, the workflow moves to step (3) for the operation of *rubber sheeting*; otherwise, it proceeds to step (4). The step of *rubber sheeting* is within a box with dotted boundary in Figure 2, because it may not be performed if there are insufficient control points. In step (4), the workflow will perform *similarity calculation* based on suitable similarity measures depending on how it reaches this step. If the workflow reaches step (4) without *rubber sheeting*, similarity measures that require correct spatial coordinates (e.g., spatial distances) will not be used, and the workflow will focus on topological relations that still hold without proper spatial references. If the workflow reaches step (4) after *rubber sheeting*, it will compute the spatial distances, topological relations, and approximate topological relations among the geographic entities, and calculate their corresponding similarity scores. Finally, step (5) combines the computed similarity scores to perform *alignment classification* and outputs the aligned entity pairs. It is worth noting that the alignment is based on the entities with the same geometry type (e.g., polygon to polygon and line to line), and the aligned entity pairs in different geometry types are combined into one set in the final result. This workflow is general, and each of the steps from (2) to (5) can be implemented by a number of possible methods. In the following, we discuss these methods.

### 3.3 Implementation methods
#### 3.3.1 Textual label alignment

When there is a lack of georeference information, we can first match entities based on their textual labels, and then identify control points based on the matched entities to further align the rest of the entities. This is similar to how we humans identify matching entities when two maps are in different orientations or scales: we first try to find an initial set of matching entities and use them as "landmarks" to align the rest. Textual label is an important type of information that we often use to find the initial set of matching entities on two maps. Here, the textual labels are already extracted and assigned to their corresponding geographic entities, and there exist methods for doing so, such as the one from Chiang and Knoblock (2012) which is based on the orientations of textual labels and their distances to corresponding geographic entities. We consider nine possible methods for the step of textual label alignment.

- *str*: This is a naive string-matching method that directly compares two place names and determines whether they are the same or not. This method is case sensitive.
- *str + caseless*: This method relaxes the previous method by ignoring upper and lower cases in the textual labels.
- *str + caseless - punc*: This method removes any punctuations in textual labels before comparison, such as commas and periods. This is useful when the textual label from one map is "3rd Av." while the label from the other map is "3rd Av".
- *str + caseless - punc - non-core*: This method further removes the *non-core terms* from place names before making a comparison. Terms in a place name can be classified into *core* and *non-core terms* (Kaffes et al. 2019). Core terms are those that represent the distinguishable toponyms (e.g., "Delaware" in "Delaware Street"), while non-core terms are usually used to describe categorical information, such as "Road" and "Avenue".

- *str + caseless - punc - non-core + domain*: This method adds further domain knowledge for enhancing the alignment of textual labels. One type of domain knowledge that we have added in this work is the equivalence between place names that are written in words and those written in numbers. For example, "3rd" is the same as "third" in place names, so that "3rd Av" will be matched to "Third Av".
- *Santos et al. (2018b)*: This is a machine learning model proposed by Santos et al. (2018b) for matching toponyms. This model computes 13 string similarity scores using a variety of measures, such as the Levenshtein edit distance, the Jaro-Winkler metric, the cosine similarity between the character n-grams of the names, and others (such as the Monge-Elkan similarity (Monge and Elkan 1996)). These similarity scores are then combined via a random forest model which determines whether two toponyms are a match. We obtain a trained model using the code and datasets shared by the authors on GitHub[3].
- *Santos et al. (2018a)*: This is a deep learning model proposed by Santos et al. (2018a) for matching toponyms. The authors first transformed the toponyms to be matched into one-hot vector representations. With one-hot vectors as the input, a recurrent neural network (RNN) model was designed to generate real-valued vector representations for toponyms. Toponyms are finally classified into matching or not based on these vectors. We obtain a trained model using their shared code and datasets on GitHub[3].
- *Acheson et al. (2019)*: This is a machine learning model proposed by Acheson et al. (2019) for toponym matching. Similar to Santos et al. (2018b), this method trains a random forest model that integrates multiple string similarity scores for comparing toponyms. However, they used only six string similarity measures, which are Levenshtein distance, Levenshtein distance with comma removed, minimum of the first two computed Levenshtein distance, the normalized Levenshtein-Damerau distance, the Jaro similarity, and the Jaro-Winkler similarity. We obtain a trained model using the code and datasets shared by the authors on GitHub[4].
- *ensemble learning*: This method combines the outputs of the three machine learning models above. Two place names are considered as a matching pair if all three models consider them as a match. We use this strict consensus requirement rather than e.g., a majority voting approach, because the aligned entities with textual labels will be used for finding control points for rubber sheeting and a stricter consensus requirement can help ensure the correctness of the aligned entities and improve the rubber sheeting result.

*3.3.2 Rubber sheeting*

Textual label alignment makes it possible to identify an initial set of matching entities from two historical maps. These entities can be used for finding control points to perform rubber sheeting, which can help overlay the two sets of entities by adjusting their spatial positions. While control points are often identified in a manual or semi-automatic manner, some automatic methods have been developed. For example, Song et al. (2008) extracted road intersections from vector road maps and raster images, and used those intersections as control points to perform rubber sheeting. Here, we combine two methods to automatically identify control points from the initial set of entities aligned by their textual labels. The first approach is similar to the one by Song et al. (2008), in which we first extract the intersections of all line features (e.g., roads and rivers), and two intersections are considered as a control point pair if the names of their line features match. Second, if the two entities matched by their names are both point features (e.g., benchmarks), their locations are directly used as control points.

---

[3] https://github.com/ruipds/Toponym-Matching
[4] https://github.com/eacheson/machine-learning-gazetteer-matching

While a combination of the two methods above can identify a set of control points, some of them can be incorrect due to the possible errors from the textual label alignment. For example, two roads may be mistakenly aligned because of their similar place names, and consequently, the control points extracted based on the intersections related to the two roads will be incorrect. Since the quality of the control points can directly affect the quality of the rubber sheeting result, we propose a three-step approach to automatically filtering out potentially incorrect control points. First, we perform an initial rubber sheeting on the two sets of entities based on all found control points. Second, we calculate the Euclidean distances for all the control point pairs based on the rubber sheeted result. The mean and standard deviation statistics of the distances are calculated, and the control point pairs whose distances are two standard deviations away from the mean are considered as potentially incorrect control points and are removed. Third, the two sets of entities are rubber sheeted again using only the control points that have passed the test. We implement the operation of rubber sheeting using the function of *transform features* from ArcPy. A minimum of three control points is required.

### *3.3.3 Similarity calculation*

This step measures the similarity between any two entities with the same geometry type based on a variety of measures. We consider two different situations when the workflow reaches this step. In the first situation, the two sets of entities are not under the same spatial reference. This happens when the historical maps do not have georeferences and cannot be rubber sheeted (e.g., due to insufficient control points). In such a case, we focus on the topological relations between entities that still hold. Specifically, we use a method based on the immediate nearby neighbors (INNs) of an entity to measure entity similarity. We consider an INN as an entity that is immediately next to the target entity. As illustrated in Figure 3, $L_1$ is an INN of the target entity $P_1$, while $P_2$ is not an INN of $P_1$, because $P_2$ is blocked by $L_1$ and thus is not immediately next to $P_1$. Computationally, the INN relation of two entities is determined in the following steps. First, the nearest points of two entities are identified. This can be implemented using the *nearest_points* function from the Python *shapely* library. Second, straight lines (e.g., $S_1$ and $S_2$) are automatically generated based on the identified nearest points. Third, if the generated line does not intersect with any other entities, then these two entities are INNs; otherwise, they are not. In the example shown in Figure 3, $P_1$ and $L_1$ are INNs because the generated line $S_1$ does not intersect with other entities; by contrast, $P_1$ and $P_2$ are not INNs because the generated line $S_2$ intersects with another entity $L_1$. With these steps, we can identify the INNs for each entity on a map. Mathematically, the INNs of a target entity $e_i$ are represented as a set: $INN_i = \{inn_{i1}, inn_{i2}, inn_{i3}, ...\}$. Note that each INN entity is identified by their unique IDs when these entities are extracted from historical maps. Given two entities $e_i$ and $e_j$, their INN-based similarity is calculated using the Jaccard index (Equation (1)):

$$J(e_i, e_j) = \frac{|INN_i \cap INN_j|}{|INN_i \cup INN_j|} \qquad (1)$$

where $|INN_i \cap INN_j|$ is the number of the entities in the two sets $INN_i$ and $INN_j$ that have already been aligned. For example, if $inn_{i1}$ is aligned to $inn_{j3}$ and $inn_{i2}$ is aligned to $inn_{j1}$, then $|INN_i \cap INN_j|$ is 4. $|INN_i \cup INN_j|$ is the total number of entities in these two sets. Initially, $INN_i \cap INN_j$ will contain only the entities that are aligned based on their textual labels in step (2) of the workflow, and this intersection will be expanded iteratively as new entity alignments are identified based on INNs. Our workflow keeps track of the entities that have already been aligned as it proceeds.

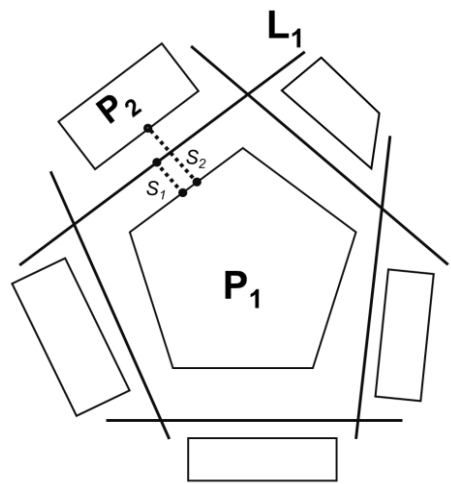

Figure 3. An illustration of finding the INNs of an entity.

The second situation is when the entities from the two maps can be put under the same spatial reference. This happens either when the entities directly come with georeferences or when the entities can be rubber sheeted. A number of similarity measures can be used in this situation:

- *spatial distances*: Since entities are under the same spatial reference, we can use spatial distance to measure their similarity. Four distance measures are used here, which are Euclidean Distance based on Centroids (EDC), shortest Euclidean Distance based on Vertices (EDV), Hausdorff Distance based on Vertices (HDV), and Euclidean Distance of the Nearest Points (EDNP). Learning from the map conflation literature (Rosen and Saalfeld 1985), we also calculate the intersecting angle for line features (e.g., roads), and combine it with their spatial distances to further measure their similarity. These distance measures have different effectiveness for entities in different geometry types (e.g., lines and polygons), and one could extend this workflow by using a different set of distance measures for each type of entities. In addition, distance measures other than the four can also be used. For example, Fréchet distance can be used for measuring the distance between two curved lines. A comprehensive survey of distance and other similarity measures were provided by Xavier et al. (2016).
- *topological relations*: Topological relations, such as the INN relation discussed previously, can also be used to measure the similarity between entities when they are put under the same spatial reference.
- *approximate topological relations*: Approximate topological relations refer to the topological relations between entities based on their broad boundaries (Clementini and Di Felice 1997). Three types of containers have been used to construct the broad boundaries of entities, which are buffer zones, minimum bounding rectangles (MBRs), and convex hulls. Here, we use buffer zones to generate broad boundaries since they better maintain the geometric shape of entities. Approximate topological relations are then computed based on the buffer zones serving as broad boundaries. Particularly, we compute the *approximately within* relation using the same approach proposed by Regalia et al. (2019): the 0.05 quantile of the entity distances is used as the buffer distance to generate the broad boundary of an entity; two entities have the relation of *approximately within* when the intersection area is greater than or equal to 80% of the area of the smaller one. It is worth noting that the approximate topological relation is calculated for entities in *different* maps, while the INN relation is calculated for entities in the *same* map.

### 3.3.4 Alignment classification

The calculated similarities from the previous step enable the alignment of the geographic entities from different historical maps. In this step, we perform alignment classification, i.e., determining whether two entities can be aligned, by combining these similarity measures using multiple possible approaches.

- *topo:* This approach is mainly used for the situation when the two sets of entities cannot be put under the same spatial reference. The alignment is performed based on the initial set of the matched entities from the step of textual label alignment, and leverages the topological relation, i.e., the INNs of an entity. Particularly, this approach is performed in three steps, as illustrated in Figure 4. In step (1), the initial set of matched entities from the textual label alignment is obtained. For example, the entities in the same colors in step (1) of Figure 4 represent the initially matched entities. In step (2), new entity alignments are identified by checking the INNs of each entity pair, and two entities are considered as a matching pair if all of their INNs match, i.e., their INN-based similarity equals to one. For example, the two polygon entities in red in step (2) of Figure 4 will be identified as a new alignment, since all their INNs match. In step (3), the newly aligned entities (e.g., the two red polygon entities) are added to the set of matched entities to update the INN-based similarity of entity pairs, and steps (2) and (3) are iterated until no new alignment is found.

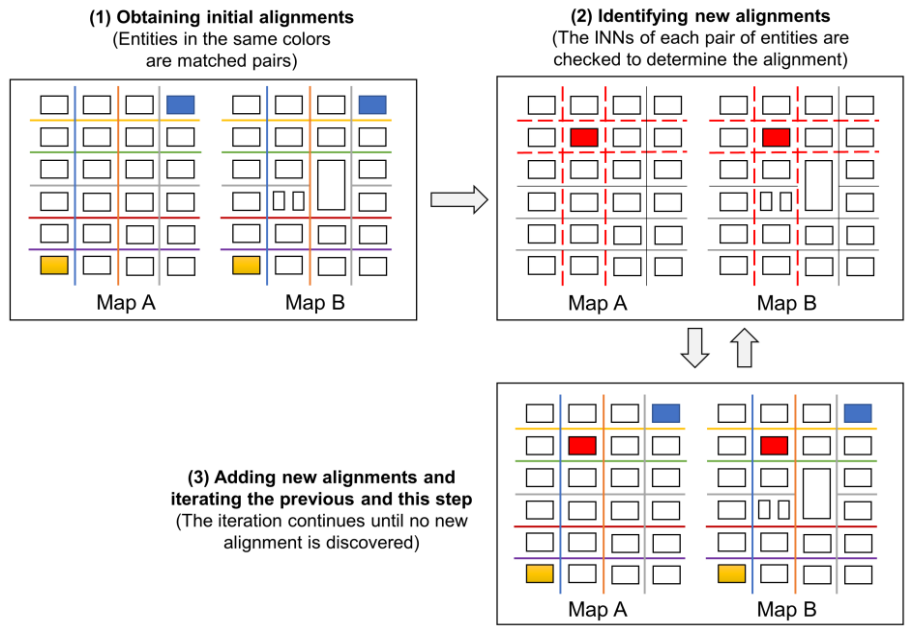

Figure 4. The iterative process of aligning geographic entities based on their INNs.

- *dist*: This and the following approaches are used for the situation when the two sets of

entities are put under the same spatial reference. Since two maps could cover different geographic extents, we perform alignment for only the entities in the overlapping area. This approach uses the four types of spatial distances discussed previously to determine the alignment, which are EDC, EDV, HDV, and EDNP. The distances between one target entity in the first map and all entities with the same geometry type in the second map are calculated, and the target entity is aligned to the entity in the second map that has the smallest distance to it. However, multiple entities in the second map could have the same distance to the target entity in the first map. When that happens, all the entities with a tied distance are removed, if distance is the only measure used for determining the alignment. This is to ensure that the output of the workflow is valid, i.e., one entity from the first map is aligned to at the most one entity in the second map. An alternative approach is to allow the target entity to be aligned to one entity randomly selected from the tied entities. However, that alternative approach is likely to produce alignment errors. When spatial distance is combined with other measures to determine alignment (such as in some of the following approaches), all entities with a tied distance are kept since they can be filtered out by the other measures. For two lines to be aligned, we also require their intersecting angle to be within 45° following the literature (Rosen and Saalfeld 1985).

- *approx*: This approach performs alignment classification based on only approximate topological relations. Two entities are considered as a match if they satisfy the relation of *approximately within*. When more than one entity from the second map have the relation of *approximately within* with target entity, the entity which has the largest overlapping ratio is aligned to the target entity. The overlapping ratio is computed by dividing the intersection area of the two broad boundaries using the area of the smaller boundary.
- *dist + topo*: This approach first generates initial alignments based on the distance approach, which are then refined by the INNs of the aligned entities. If there exists no single matching pair in the INNs of two initially aligned entities, this alignment will be considered as incorrect and will be removed.
- *dist + approx*: Similar to the previous approach, this approach first generates initial alignments based on the distance approach, which are then refined based on the relation of *approximately within*.
- *approx + topo*: This approach first generates initial alignments based on the relation of *approximately within* and then refines the result based on the INNs of each entity.
- *dist + topo + approx*: This final approach combines spatial distances, topological relations, and approximate topological relations for alignment classification.

## 4 Experiments

Since multiple possible methods can be used for completing each step of the proposed workflow, which methods can provide better performances? In this section, we evaluate and understand the performances of these methods and the overall workflow. Two datasets of historical maps are used for the experiments. In the following, we first describe the datasets and the experiment procedure and then discuss the results.

### 4.1 Datasets

Two datasets of historical maps are used in our experiments. The first dataset contains three historical Sanborn Fire Insurance maps covering the city of Buffalo, NY, USA in the years of 1889, 1899, and 1925 (Figure 5). The second dataset contains three historical maps covering the University at Buffalo South Campus (UBSC) in the years of 1966, 1982, and 1990 (Figure 6). Both sets of maps were obtained from the library of the University at buffalo in the form of scanned map images.

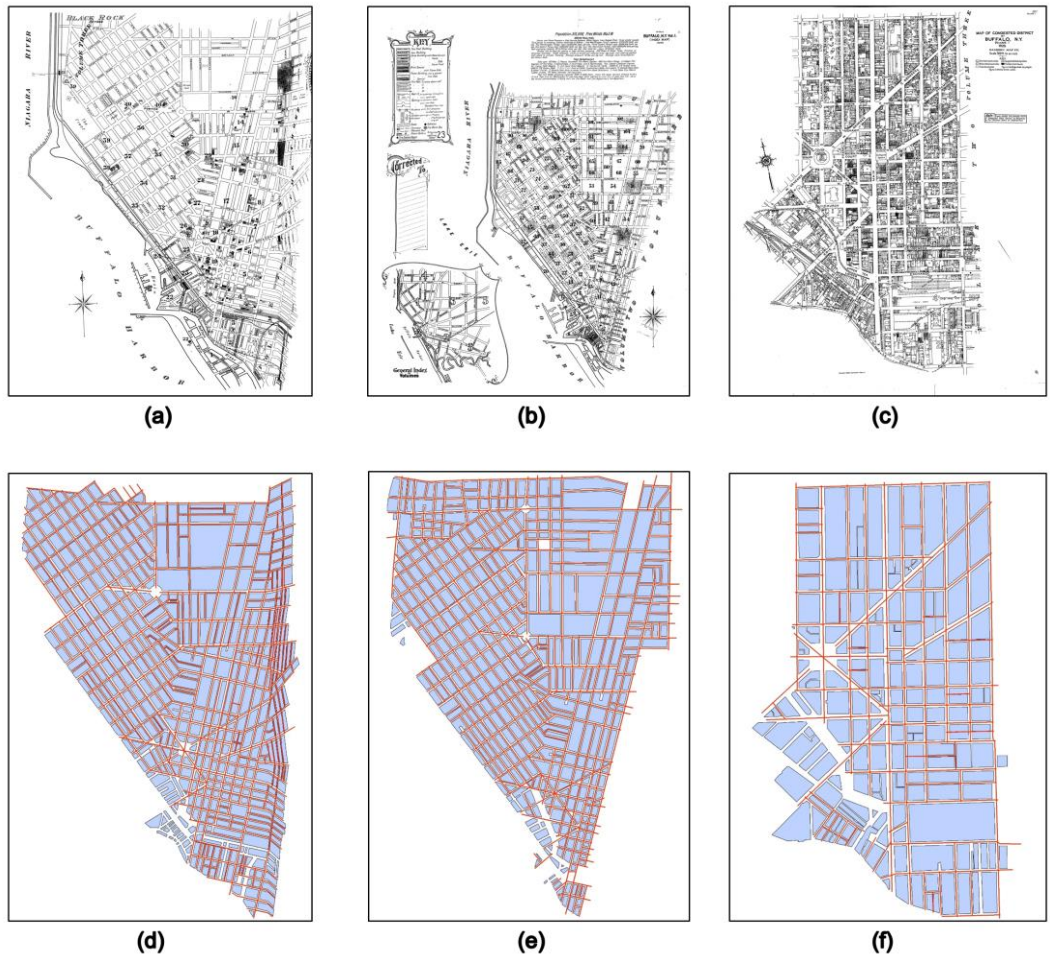

Figure 5. Three Sanborn Fire Insurance maps covering the city of Buffalo, NY, USA; (a) (b) (c) are scanned maps in 1889, 1899, and 1925; (d) (e) (f) are vectorized entities.

Given our focus on entity alignment, we extract entities from the two sets of historical maps using a combination of the software tool ArcScan and manual editing. Only the information presented on the maps are passed onto the extracted entities. We show these vector entities in the lower sections of Figure 5 and Figure 6. As can be seen, these map images can be in different orientations (e.g., the three UBSC maps) depending on how the maps were made or scanned, and can cover different geographic extents (e.g., the Buffalo maps). A polyline layer and a polygon layer are created for each map. The polyline layer contains roads and rivers, while the polygon layer contains buildings and city blocks. Table 1 summarizes the numbers of different types of entities in the two datasets. Fewer than 30% entities are associated with textual labels, and both datasets do not have georeferences. These two datasets make the task of aligning geographic entities challenging, while reflecting some of the difficulties that one may encounter in a real-world application.

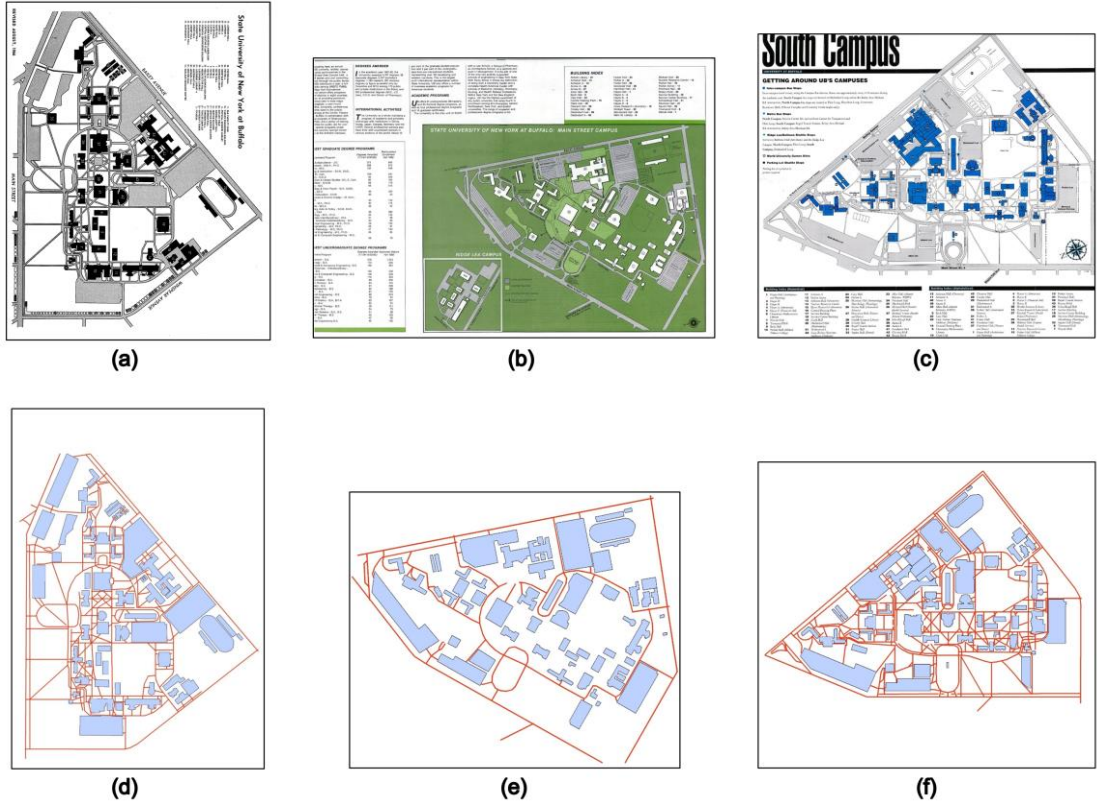

Figure 6. The historical maps of the University at Buffalo South Campus; (a) (b) (c) are scanned maps in the years of 1966, 1982, and 1990; (d) (e) (f) are the vectorized entities.

Table 1. Numbers of different types of entities in the experimental datasets.

| Dataset | Categories of entity | Map years | | | |
|---|---|---|---|---|---|
| | | *1889* | *1899* | *1925* | *In total* |
| Buffalo dataset | Entities with polyline geometry | 203 | 196 | 74 | 473 |
| | Entities with polygon geometry | 548 | 449 | 190 | 1187 |
| | Total number of entities | 751 | 645 | 264 | 1660 |
| | *Entities with textual labels* | 163 (21%) | 176 (27%) | 66 (25%) | 405 (24%) |
| Dataset | Categories of entity | Map years | | | |
| | | *1966* | *1982* | *1990* | *In total* |
| UBSC dataset | Entities with polyline geometry | 211 | 42 | 252 | 505 |
| | Entities with polygon geometry | 63 | 57 | 58 | 178 |
| | Total number of entities | 274 | 99 | 310 | 683 |
| | *Entities with textual labels* | 47 (17%) | 66 (67%) | 66 (21%) | 179 (26%) |

To evaluate the performances of the proposed workflow and methods, we manually identify matching entities from two maps in the same dataset that are in two consecutive years (e.g., the maps of 1889 and 1899 in the Buffalo dataset, or the maps of 1966 and 1982 in the UBSC dataset). The matching entities are identified via visual inspection, and are used as the ground truth for the evaluation experiments. In total, there are 584 matching entity pairs in the Buffalo dataset, and 163 matching entity pairs in the UBSC dataset. The entities are in different sizes and orientations. Table 2 summarizes these matching entity pairs in the ground truth.

Table 2. Numbers of the matching entity pairs in the ground truth.

| Dataset | Category of entity alignments | Map pairs | |
|---|---|---|---|
| | | *1889-1899* | *1899-1925* |
| Buffalo dataset | Matching entity pairs with polyline geometry | 145 | 49 |
| | Matching entity pairs with polygon geometry | 323 | 67 |
| | Total number of alignments | 468 | 116 |

| Dataset | Category of entity alignments | Map pairs | |
|---|---|---|---|
| | | *1966-1982* | *1982-1990* |
| UBSC dataset | Matching entity pairs with polyline geometry | 29 | 31 |
| | Matching entity pairs with polygon geometry | 50 | 53 |
| | Total number of alignments | 79 | 84 |

## 4.2 Experiment procedure and results

With the two datasets and ground truth prepared, we start to evaluate the performances of the possible methods and the workflow. Three metrics from information retrieval, namely *precision*, *recall*, and *F-score*, are used for the evaluation experiments. *Precision* is the percentage of correctly identified alignments among all the alignments identified by a method, which include both correct and incorrect alignments. *Recall* is the percentage of correctly identified alignmetns among all the correct alignments in the ground truth. *F-score* is the harmonic mean of *precision* and *recall*. These three metrics are defined in Equations (2)-(4). Note that the alignments in the ground truth are encoded in the form of $[e_{Ai}, e_{Bj}]$, where $e_{Ai}$ is the *ith*

$$\text{precision} = \frac{|correctly\ identified\ alignments|}{|all\ identified\ alignments|} \quad (2)$$

$$\text{recall} = \frac{|correctly\ identified\ alignments|}{|all\ correct\ alignments|} \quad (3)$$

$$\text{F} - \text{score} = \frac{2 \times precision \times recall}{precision + recall} \quad (4)$$

entity on map $A$ (represented by its entity ID) and $e_{Bj}$ is the *jth* entity on map $B$. The alignments identified by a method are encoded in the same form, represented as $[e_{Ai\prime}, e_{Bj\prime}]$. By comparing the identified alignments with those in the ground truth, we can obtain the number of correctly identified alignments.

The experiments are conducted following the individual steps of our workflow, in which we first test the different methods for textual label alginment, then perform rubber sheeting and similarity calculation, and finally examine the performances of the multiple alignment classificaiton approaches. Since both datasets do not have georeferencing information, we start by aligning entities based on the available textual labels. Specifically, we apply the nine textual label alignment methods discussed in Section 3.3.1 to the available place names in the two datasets, and the results are reported in Table 3 and Table 4. Please note that the *precision*, *recall*, and *F-score* in Table 3 and Table 4 are calculated based on only the entities that have textual labels rather than all the entities in the data, since the methods are not applicable to entities without textual labels.

Table 3. Performances of the nine textual label alignment methods on the Buffalo dataset.

|  | 1889-1899 | | | 1899-1925 | | |
|---|---|---|---|---|---|---|
| *Method* | *Precision* | *Recall* | *F-score* | *Precision* | *Recall* | *F-score* |
| str | 0.8652 | 0.6311 | 0.7299 | **0.8571** | 0.6977 | **0.7692** |
| str + caseless | 0.8652 | 0.6311 | 0.7299 | **0.8571** | 0.6977 | **0.7692** |
| str + caseless - punc | **0.8750** | 0.6885 | 0.7706 | **0.8571** | 0.6977 | **0.7692** |
| str + caseless - punc - non-core | **0.8750** | 0.6885 | 0.7706 | **0.8571** | 0.6977 | **0.7692** |
| str + caseless - punc - none-core + domain | 0.8624 | 0.7705 | **0.8139** | **0.8571** | 0.6977 | **0.7692** |
| Santos et al. (2018b) | 0.0112 | 0.8770 | 0.0222 | 0.0121 | **0.9651** | 0.0239 |
| Santos et al. (2018a) | 0.0251 | **0.9098** | 0.0484 | 0.0252 | 0.8837 | 0.0485 |
| Acheson et al. (2019) | 0.5056 | 0.7377 | 0.6000 | 0.4769 | 0.7209 | 0.5741 |
| ensemble learning | 0.7377 | 0.7377 | 0.7377 | 0.6327 | 0.7209 | 0.6739 |

Table 4. Performances of the nine textual label alignment methods on the UBSC dataset.

|  | 1966-1982 | | | 1982-1990 | | |
|---|---|---|---|---|---|---|
| *Method* | *Precision* | *Recall* | *F-score* | *Precision* | *Recall* | *F-score* |
| str | **1.0000** | 0.0465 | 0.0889 | **1.0000** | 0.3906 | 0.5618 |
| str + caseless | 0.8846 | 0.5349 | **0.6667** | **1.0000** | 0.6094 | 0.7573 |
| str + caseless - punc | 0.8846 | 0.5349 | **0.6667** | **1.0000** | 0.6250 | **0.7692** |
| str + caseless - punc - non-core | 0.8846 | 0.5349 | **0.6667** | **1.0000** | 0.6250 | **0.7692** |
| str + caseless - punc - non-core + domain | 0.8846 | 0.5349 | **0.6667** | **1.0000** | 0.6250 | **0.7692** |
| Santos et al. (2018b) | 0.0453 | 0.5349 | 0.0834 | 0.1469 | **0.8359** | 0.2499 |
| Santos et al. (2018a) | 0.0940 | **0.5698** | 0.1559 | 0.1766 | 0.7109 | 0.2817 |
| Acheson et al. (2019) | 0.1818 | 0.0465 | 0.0741 | 0.7429 | 0.4062 | 0.5253 |
| ensemble learning | **1.0000** | 0.0465 | 0.0889 | 0.9286 | 0.4062 | 0.5652 |

In these two tables, the former five are simple string-matching methods while the latter four are machine learning based methods. As can be seen, simple string-matching methods generally outperform machine learning based methods. This result is surprising, as we initially expected that the more advanced machine learning models integrating multiple string similarity measures would have better performances than naive string-matching methods. Particularly, in the results of the Buffalo dataset, the precisions of Santos et al. (2018b) and Santos et al. (2018a) are extremely low, which suggest that many incorrect alignments are included. To understand why, we look into the experiment results as well as the models themselves. We find that these two machine learning methods make matching decisions based on the thresholds derived from the training data. The training data are toponyms and their alternatives in the GeoNames gazetteer. Since the training data consider a place name (e.g., "*Lisboa*") and its alternative name (e.g., "*Olissipona*") as a match, the learned thresholds are too low for the place names in the Buffalo dataset, which result in many false positives and thus low precisions. The machine learning model by Acheson et al. (2019) is trained on a different dataset, and the learned threshold seems to be more suitable for matching the place names in the Buffalo dataset. However, we see largely different performances of these machine learning models on the UBSC dataset. Many place names in the UBSC dataset are in different upper and lower cases, and therefore their string similarity scores are lower than those in the Buffalo dataset. Since the same thresholds are used by the machine learning models, they show different performances in identifying place name matches in the UBSC dataset.

These experiment results suggest that machine learning including deep learning models are highly sensitive to their thresholds obtained from their particular training data. Due to the likely difference between the training data and the data from the actual application, pre-trained machine learning models may not perform well and should ideally be retrained or fine-tuned using the data from the actual application. However, there is usually a lack of labeled training data in the context of aligning entities from historical maps, which limits the application of machine learning models. By contrast, simple string-matching methods can provide overall good performances without requiring labeled training data.

Based on the experiment results in Table 3 and Table 4, we choose the third textual alignment method, namely *str + caseless - punc*, for performing the following experiments. We choose this method for three reasons. First, *str + caseless - punc* is among the string-matching methods which are more robust than the other four machine learning methods. Second, this method does not require external information, such as core and non-core terms or digital numbers and their word equivalents, and therefore has better generalizability. Third and most importantly, this method achieves fairly high scores for both precision and recall. We need high precision since the textual alignment result will be used for determining the control points for rubber sheeting; we also need high recall since a very low recall can lead to insufficient control points (note that when a large number of control points exist, precision is more important than recall).

Using the alignment results from *str + caseless - punc*, we identify control points and perform rubber sheeting to overlay the entities in the datasets. We use the three-step method discussed previously to identify high-quality control points. The numbers of the identified control points for these four map pairs are 189, 56, 1, and 12 respectively. As a result, we can perform rubber sheeting for the Buffalo maps 1889-1899 and 1899-1925 as well as the UBSC maps 1982-1990. We cannot perform rubber sheeting for the UBSC maps 1966-1982 since there is only one control point. The automatically rubber sheeted results of Buffalo maps 1889-1899 and 1899-1925, and UBSC maps 1982-1990 are shown in Figure 7.

With the rubber sheeted entities, similarity scores are computed between entities based on the spatial distances, topological relations, and approximate topological relations among entities. We then examine the performances of the seven alignment classification approaches for combining these similarity scores and aligning the entities. Since the spatial distance between two entities can be measured in different ways, we first look into the performances of using spatial distance alone for alignment classification, and test the four different spatial distance measurements, namely EDC, EDV, HDV, and EDNP. The results are reported in Table 5 and Table 6. Note that the distance-based approach cannot be applied to the UBSC map pair 1966-1982, since these two maps are not rubber sheeted due to insufficient control points. In addition, the *precision, recall, and F-score* reported in Table 5, Table 6, and the following tables are based on all the entities in the datasets rather than only those with textual labels.

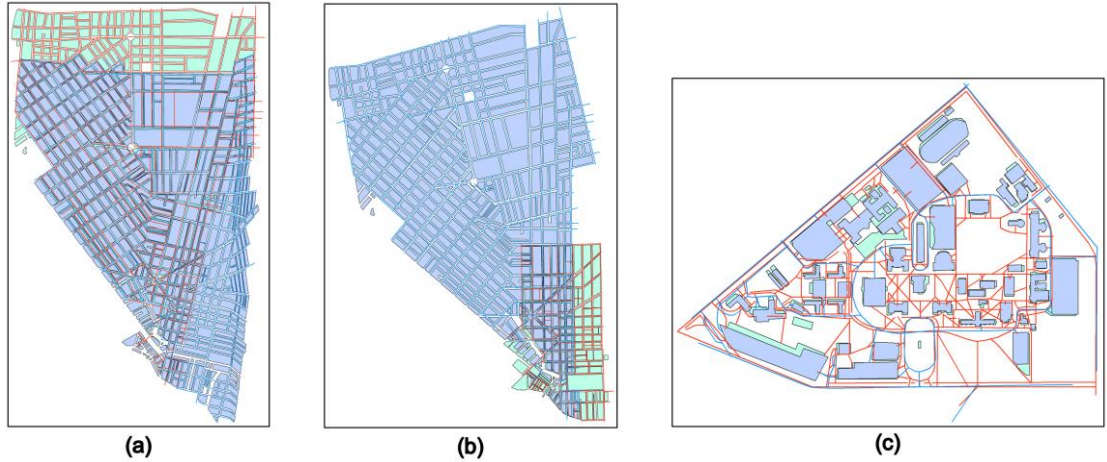

Figure 7. The automatically rubber sheeted results of the Buffalo and UBSC datasets; (a) (b) are the results of the Buffalo maps 1889 and 1899, and 1899 and 1925; (c) is the result of the UBSC maps 1982 and 1990.

Table 5. Performances of different distance measures for alignment on the Buffalo dataset.

|  | *1889-1899* | | | *1899-1925* | | |
| --- | --- | --- | --- | --- | --- | --- |
| *Distance Metric* | *Precision* | *Recall* | *F-score* | *Precision* | *Recall* | *F-score* |
| EDC | 0.7316 | **0.9551** | **0.8285** | 0.6485 | 0.9224 | 0.7616 |
| EDV | 0.6574 | 0.8611 | 0.7456 | 0.6265 | 0.8966 | 0.7376 |
| HDV | 0.7217 | 0.9530 | 0.8214 | 0.6585 | **0.9310** | 0.7714 |
| EDNP | **0.7455** | 0.8761 | 0.8055 | **0.7754** | 0.9224 | **0.8425** |

Table 6. Performances of different distance measures for alignment on the UBSC dataset.

|  | *1966-1982* | | | *1982-1990* | | |
| --- | --- | --- | --- | --- | --- | --- |
| *Distance Metric* | *Precision* | *Recall* | *F-score* | *Precision* | *Recall* | *F-score* |
| EDC | - | - | - | 0.8791 | 0.9524 | 0.9143 |
| EDV | - | - | - | 0.8132 | 0.8810 | 0.8457 |
| HDV | - | - | - | **0.8830** | **0.9881** | **0.9326** |
| EDNP | - | - | - | 0.8276 | 0.8571 | 0.8421 |

As can be seen from these two tables, different spatial distance measures can result in different entity alignment results. Meanwhile, all four distance measurements have achieved fair performances, and there is no one single method that dominates all the others on the two datasets of historical maps. We eventually choose HDV (Hausdorff Distance based on Vertices) as our method for measuring the spatial distance between two entities, and use it in the following alignment experiments. We choose HDV because it achieves high recalls which can help retain correct alignments in the initial result that can be further refined by the other measures based on topological relations and approximate topological relations.

Using HDV as the spatial distance measure, we systematically examine the performances of the seven alignment classification approaches on the two datasets. Table 7 and Table 8 summarize their performances. In both tables, the first row, *topo,* represents the performances of using topological relations to align entities without rubber sheeting. Thus, we separate this row from the following rows using a thicker line to differentiate the situations *without rubber sheeting* (the first row) and *with rubber sheeting* (the following rows). In addition, we provide bar charts in Figure 8 to illustrate the performances of the seven alignment approaches.

Table 7. Performances of the seven alignment approaches on the Buffalo dataset.

|  | *1889-1899* | | | *1899-1925* | | |
|---|---|---|---|---|---|---|
| *Approach* | *Precision* | *Recall* | *F-score* | *Precision* | *Recall* | *F-score* |
| topo | **0.9122** | 0.2885 | 0.4383 | **0.8667** | 0.3362 | 0.4845 |
| approx | 0.8047 | **0.9594** | 0.8752 | 0.7152 | 0.9310 | 0.8090 |
| approx + topo | 0.8156 | 0.9359 | 0.8716 | 0.7315 | **0.9397** | 0.8226 |
| dist | 0.7217 | 0.9530 | 0.8214 | 0.6585 | 0.9310 | 0.7714 |
| dist + topo | 0.7587 | 0.9274 | 0.8346 | 0.6792 | 0.9310 | 0.7855 |
| dist + approx | 0.8180 | 0.9509 | **0.8794** | 0.7681 | 0.9138 | 0.8346 |
| dist + topo + approx | 0.8295 | 0.9252 | 0.8747 | 0.7852 | 0.9138 | **0.8446** |

Table 8. Performances of the seven alignment approaches on the UBSC dataset.

|  | *1966-1982* | | | *1982-1990* | | |
|---|---|---|---|---|---|---|
| *Approach* | *Precision* | *Recall* | *F-score* | *Precision* | *Recall* | *F-score* |
| topo | **0.8846** | **0.2911** | **0.4381** | **1.0000** | 0.4762 | 0.6452 |
| approx | - | - | - | 0.8191 | 0.9167 | 0.8652 |
| approx + topo | - | - | - | 0.8500 | 0.8095 | 0.8293 |
| dist | - | - | - | 0.8830 | **0.9881** | 0.9326 |
| dist + topo | - | - | - | 0.9221 | 0.8452 | 0.8820 |
| dist + approx | - | - | - | 0.9432 | **0.9881** | **0.9651** |
| dist + topo + approx | - | - | - | 0.9726 | 0.8452 | 0.9045 |

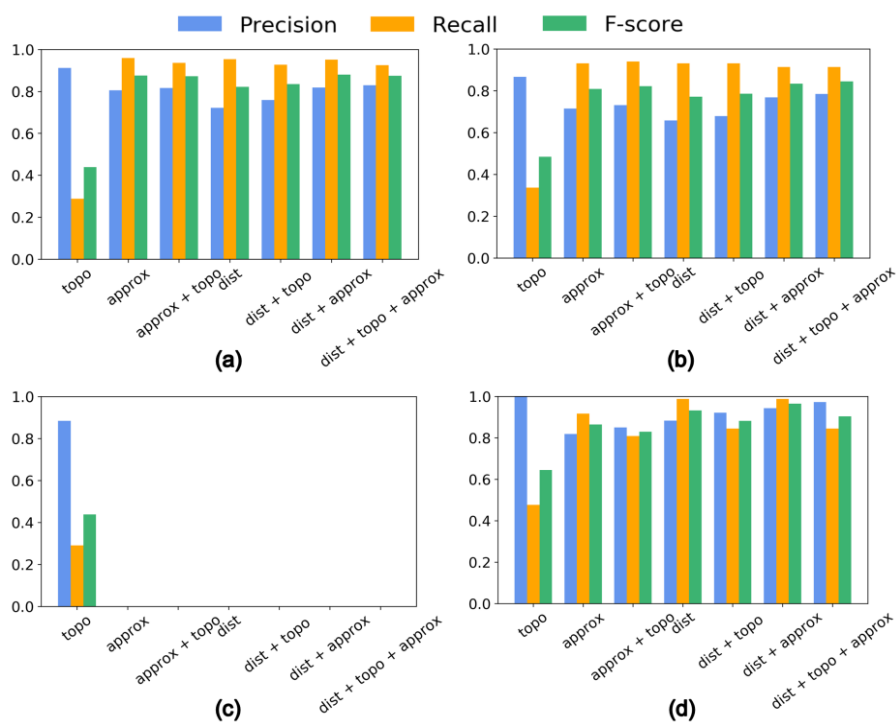

Figure 8. Performances of the seven alignment approaches illustrated as bar charts: (a) and (b) are based on the Buffalo dataset; (c) and (d) are based on the UBSC dataset.

## 4.3 Discussion
### *4.3.1 Result interpretation*
Table 7 and Table 8 show the performances of the seven alignment approaches, as well as the final performances of the proposed workflow given different implementations. As can be seen, the approach of *topo* achieves very high precisions, which suggests that most of its identified alignments are correct. This can be attributed to the strict alignment requirement of this approach, namely two entities are considered as a match only if all of their INNs are matched. This result is encouraging as it shows that we can still obtain high-precision alignments even when two maps cannot be put under the same spatial reference. A drawback of using *topo* alone is its low recalls. Since this approach largely depends on the initial set of alignments based on the textual labels of entities, it cannot identify many more matching pairs if there are only a small number of aligned entities in the initial set.

When two sets of entities can be put under the same spatial reference, alignment methods that rely on spatial coordinates can be used (starting from the second row in Tables 7 and 8). We discuss these results by organizing them into three groups: approaches based on one measure (i.e., the second and fourth rows), two measures (i.e., the third, fifth, and sixth rows), and all three measures (i.e., the seventh row).

The approaches in the first group use either *approx* or *dist* alone to align the entities from two historical maps. They achieve very high recalls (from 0.9167 to 0.9594 for *approx* and from 0.9310 to 0.9881 for *dist*), indicating that both approaches can find most of the correct alignments. Their precisions, however, are lower than their recalls (from 0.7152 to 0.8191 for *approx* and from 0.6585 to 0.8830 for *dist*), suggesting that about 12% to 35% of the identified alignments are incorrect. By examining the results, we find that a major reason of these errors is that these two approaches always attempt to find an alignment for a target entity, even when such an alignment does not exist. In other words, these two approaches cannot handle one-to-zero alignment. One example is illustrated in Figure 9, in which the city block in the map of

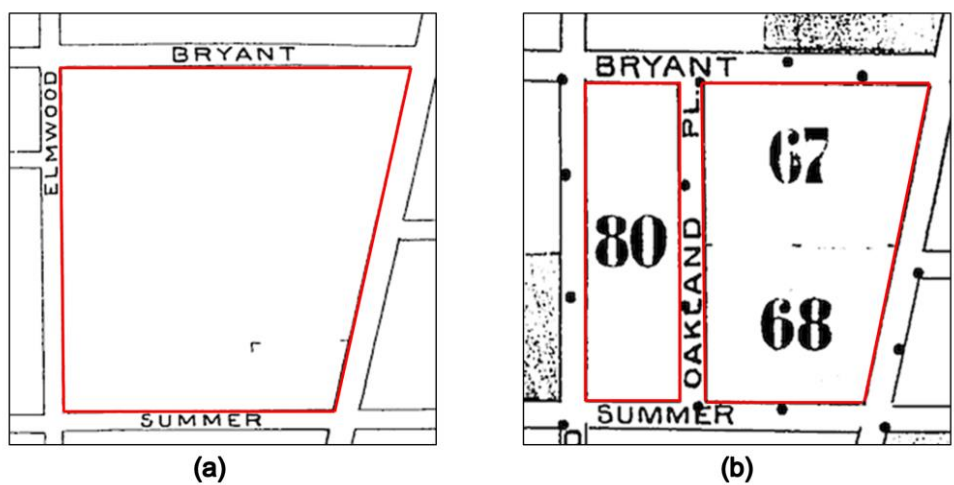

Figure 9. Illustration of an alignment error introduced by *approx* or *dist*: the city block in (a) should be aligned to neither of the two blocks in (b), but the approach of *approx* or *dist* still makes an alignment.

Buffalo in 1889 (sub figure (a)) is represented as two city blocks in the map of 1899 (sub figure (b)). Without additional domain knowledge, it is hard to know whether this is a representation difference or whether the road "Oakland Pl" was indeed built after 1889. Thus, in the ground truth, the big city block in sub figure (a) is aligned to neither of the two city blocks in sub figure (b). The approach of *approx* or *dist*, however, still makes an alignment based on the *approximately within* relation or the shortest spatial distance. For example, *approx* matches the big city block in (a) to the larger city block on the right in (b), introducing an error into the result.

The approaches in the second group, namely *approx + topo* (row three in Tables 7 and 8), *dist + topo* (row five), and *dist + approx* (row six), first use one measure to obtain an initial alignment result and then use the second measure to refine the result. For *approx + topo* and *dist + topo*, these two approaches first obtain an initial set of alignments using either *approx* or *dist*, which are then refined by *topo* based on the INNs of the entities. An initial matching pair will be removed if none of their INNs match. Compared with using *approx* or *dist* alone, *approx + topo* and *dist + topo* improve in precision, and this result indicates that the refinement based on INNs effectively removes some incorrect alignments in the initial results. Meanwhile, some correct alignments are also removed in this refinement process as shown in the decreased recalls. This is due to the lack of matching pairs in the INNs of some initially aligned entities. Particularly, there is a large decrease of recall in the maps of UBSC in 1982-1990. As shown in Figure 6 (b) and (c), these two UBSC maps have different levels of details for the roads: the map in 1990 depicts the minor roads on campus while the map in 1982 does not. These different levels of details affect the INNs of the entities and decrease the recalls consequently. The approach *dist + approx* identifies an initial set of alignments using spatial distance and then refines the result if the initial alignments also satisfy the *approximately within* relation. Compared with using *dist* alone, this approach largely increases the precision while still keeping high recalls. This result suggests that the approach *dist + approx* effectively removes the incorrect alignments that are spatially close but do not satisfy the *approximately within* relation, while keeping most correct alignments in the result.

Finally, the approach in the last group, namely *dist + topo + approx*, combines all three measures. Compared with *dist + approx* which achieves the best performance so far, *dist + topo + approx* shows an increased precision but a decreased recall. While *dist + topo + approx* is the most complex alignment approach among the seven, it does not achieve the overall best performance due to the lack of matching INNs required by *topo*.

In sum, we have conducted systematic experiments to examine and understand the performances of different methods for the individual steps of the proposed workflow. These experiments are based on two different datasets of real-world historical maps. The results show that our alignment workflow can handle the challenges of missing georeferences and the lack of textual labels in some entities. Using *str + caseless – punc* for textual label alignment, HDV for spatial distance measure, and *dist + approx* for alignment classification, the workflow achieves an average of 0.89 F-score on the two historical map datasets. The experiment results also show the pros and cons of the tested methods, such as the sensitive threshold of machine learning models for textual label alignment. We share the source code of the implemented workflow and the two historical map entity datasets for supporting future research.

### 4.3.2 Limitations of the workflow

There are several limitations of the proposed workflow. First, it relies on the availability of some textual labels on the maps to obtain an initial set of aligned entities. Accordingly, it cannot function when there is no textual label at all or only one or two labels on the maps. In those cases, one may have to first manually identify a number of matching entities to jump start the workflow. Second, this workflow can only identify one-to-one (and also one-to-zero) alignments, and will not align the same geographic features that are represented as different numbers of entities on two historical maps. As discussed before, it is difficult to distinguish these representation differences from real-world changes that indeed happened. It is worth noting that our workflow is robust to small changes in the shape and size of the geographic features to be aligned. For example, two corresponding geographic features with slightly different geometric representations on two maps will still be aligned, as long as their topological relations with other entities remain the same. Third, our current workflow focuses on identifying only matching entities which could be linked via *same-as* relations in a knowledge graph. Depending on the applications, geographic entities can have many other types of relations that can be extracted and added into a GKG.

## 5 Conclusions and future work

In this paper, we have presented an automatic workflow for aligning geographic entities extracted from different historical maps. This workflow involves a number of steps. First, it identifies matching entity pairs based on the textual labels available on some entities. Then, control points are identified using the matching entity pairs, which are used for performing rubber sheeting. After that, similarity scores are computed among entities, and finally, alignments are determined based on a combination of these similarity scores. We discuss possible methods for each of these steps, and conduct experiments on two different datasets of historical maps to systematically evaluate their performances. The experiment results demonstrate the advantages and weaknesses of different methods, and can guide the selection of methods for using this workflow.

We can build a geographic knowledge graph from historical maps using the proposed workflow for the step of aligning and linking entities. We can first generate Uniform Resource Identifiers (URIs) for the geographic entities extracted from historical maps, and add attributes to these entities, such as the year of a historical map. The proposed workflow can then be employed to find entity alignments from different maps, and Resource Description Framework (RDF) can be utilized to represent the aligned entities as triples. For example, two aligned geographic entities can be represented as *<URI$_1$> <:sameAs> <URI$_2$>,* where *URI$_1$* and *URI$_2$* are two URIs for the entities from two maps respectively and *:sameAs* is a predicate used to represent the matching relationship between the two. Depending on the targeted application of the GKG, one may want to generate additional triples capturing other relationships among the geographic entities (e.g., the *nearby* relationship between a building and a street), following the Linked Data principles (Hitzler et al. 2009, Bizer et al. 2009). These generated and linked triples form the GKG, which can be published and stored in a graph database, such as GraphDB, and encoded using a format, such as RDF/XML or JSON-LD. Finally, the SPARQL query language can be employed to query the constructed GKG (Scheider et al. 2019), through which questions such as "which buildings were constructed along Washington Street between the years of 1889 and 1925" can be answered.

Two directions can be pursued in the near future. First, in terms of methodology, we can continue to improve the alignment workflow. Given that our current workflow relies on the availability of textual labels to identify an initial set of alignments, we can explore the possibility of using the visual patterns of entities from two maps to generate initial alignments. For example, we could check the shapes of entities and their topological relations in local areas of a map using a moving window and compare them with the second map. Such an approach, however, would also be sensitive to the geometric distortion existing in some historical maps when they were scanned into images, and a combination of textual labels (if available) and visual patterns might be better. Second, in terms of applications, we can construct and implement a GKG from a collection of historical maps using selected entity extraction methods, our alignment workflow, and other related methods. We can further assess the efficacy of our alignment workflow by examining the capability of the constructed GKG in correctly answering various queries. Historical maps are important for understanding the past of a geographic area. Current and future research on integrating historical maps and GKG can help facilitate the access and use of the geographic information contained in these valuable resources.

**Data and codes availability statement**
The data and codes that support the findings of this study are available at the following link: https://doi.org/10.6084/m9.figshare.13158098.

**Acknowledgement**
The authors would like to thank the three anonymous reviewers for their constructive comments and suggestions. Kai Sun acknowledges support from the China Scholarship Council. Yingjie Hu acknowledges support from the University at Buffalo Research Foundation. Jia Song and Yunqiang Zhu acknowledge support from the Natural Science Foundation of China (No. 41771430 and 41631177) and the Strategic Priority Research Program of the Chinese Academy of Sciences (No. XDA23100100).